\documentclass[%
 reprint,
 amsmath,amssymb,
 aps,
]{revtex4-2}

\AtBeginDocument{\RenewCommandCopy\qty\SI} 

\usepackage{array}
\usepackage{xcolor,graphicx}
\usepackage{dcolumn}
\usepackage{bm}
\usepackage{wrapfig}
\usepackage{siunitx} 
\usepackage{physics}
\let\qty\SI 
\usepackage{nccmath,amsmath,lipsum}
\usepackage{gensymb}
\usepackage{comment}

\usepackage[colorlinks=true]{hyperref}
\usepackage{soul}
\usepackage{ulem}
\usepackage{colortbl}
\usepackage{appendix}

\newcommand{\revisePRL}[1]{{\color{black}#1}}
\begin{document}

\preprint{APS/123-QED}

\title{The impact of hole \textit{g}-factor anisotropy on spin-photon entanglement generation with InGaAs quantum dots}

\author{P. R. Ramesh$^{1,2}$}
\email{rameshpr@udel.edu}
\author{E. Annoni$^{1,3}$}
\author{N. Margaria$^{3}$}
\author{D. A. Fioretto$^{1,3}$}
\author{A. Pishchagin$^{3}$}
\author{M. Morassi$^{1}$}
\author{A. Lema\^itre$^{1}$}
\author{M. F. Doty$^{2}$}
\author{P. Senellart$^{1}$}
\author{L. Lanco$^{1,4}$}
\author{N. Belabas$^{1}$}
\author{S. C. Wein$^{3}$}
\author{O. Krebs$^{1}$}
\email{olivier.krebs@cnrs.fr}

\affiliation{$^{1}$Universit\'{e} Paris-Saclay, CNRS, Centre de Nanosciences et de Nanotechnologies, 91120 Palaiseau, France}
\affiliation{$^{2}$University of Delaware, Newark, DE 19716}
\affiliation{$^{3}$Quandela SAS, 91300 Massy, France}
\affiliation{$^{4}$Universit\'{e} Paris Cit\'{e}, 75013 Paris, France}

\date{\today}

\begin{abstract}
Self-assembled InGaAs/GaAs quantum dots (QDs) are of particular importance for the deterministic generation of spin-photon entanglement. One promising scheme relies on the Larmor precession of a spin in a transverse magnetic field, which is governed by the in-plane $g$-factors of the electron and valence band heavy-hole. We probe the origin of heavy-hole $g$-factor anisotropy with respect to the in-plane magnetic field direction and uncover how it impacts the entanglement generated between the spin and the photon polarization. First, using polarization-resolved photoluminescence measurements on a single QD, we determine that the impact of valence-band mixing dominates over effects due to a confinement-renormalized cubic Luttinger $q$ parameter. From this, we construct a comprehensive hole $g$-tensor model. We then use this model to simulate the concurrence and fidelity of spin-photon entanglement generation with anisotropic hole $g$-factors, which can be \revisePRL{leveraged by tuning} magnetic field angle and excitation polarization. The results demonstrate that post-growth control of the hole $g$-factor can be used to improve spin-photon cluster state generation.
\end{abstract}
\maketitle

Self-assembled semiconductor quantum dots (QDs) are important resources for photonic quantum information processing. For example, InGaAs/GaAs QDs hosting a single charge carrier are a proven deterministic source of entanglement between a spin and one or more single photons~\cite{Schwartz2016,Lee2019,Coste2023,Meng2024}. One protocol for the generation of such spin-photon entanglement uses optical excitation pulses synchronized to the coherent Larmor precession of a single resident spin around a transverse magnetic field (i.e.~perpendicular to the optical and QD growth axis $z$)~\cite{Lindner2009}. This protocol has recently been used to demonstrate on-demand generation of 4-partite entangled states with high rates, high photon indistinguishability, and promising scalability~\cite{Huet2024,Su2024}. The fidelity of the generated state is predominantly limited by the loss of resident spin coherence during the time between excitation pulses. However, the fidelity is also reduced by unwanted precession of the unpaired spin in both the optically excited and optical ground states (e.g.~the hole in the negative trion excited state, the electron in the ground state) during photon emission. This spin precession during optical emission leads to a dephasing in the target state with respect to the precisely timed sequence of excitation pulses, and thus a reduction of fidelity. To mitigate this effect for a given QD, one needs precise knowledge of the Zeeman Hamiltonian in a transverse magnetic field, namely of the electron and hole in-plane $g$-factors. This is particularly useful to determine  if there is any preferential orientation of the sample with respect to the applied magnetic field.\\
\indent For an electron in a low-dimensional epitaxial structure with nanometer-scale confinement along $z$, the in-plane $g$-factor can be considered mostly isotropic. For the hole, however, the in-plane $g$-factor is a tensor with components that are dependent on the symmetry of the confinement potential~\cite{Koudinov2004,Krizhanovskii2005,Kowalik2007,Leger2007,Kusrayev99,Kaji2017,Schwan2011,Trifonov2021a}. In self-assembled QDs, the dominant terms of this tensor are generally ascribed to a reduced $C_{2v}$ symmetry that arises when the QD shape or strain fields break roto-inversion $\bar{4}$ of the $D_{2d}$ symmetry group.
Post-growth annealing of InAs QDs has been shown to change the QD morphology~\cite{Keizer2012} and recent work suggests that strong annealing can significantly reduce the magnitude of the $C_{2v}$ perturbation ~\cite{Trifonov2021a}, which could in turn alter the $g$-factor anisotropy. \\
\indent In this Letter, we investigate the origin of anisotropic in-plane hole $g$-factors in annealed InGaAs QDs via polarization-resolved magneto-photoluminescence (PL) measurements under a varying magnetic field angle. We identify and quantify the dominant contributions to the hole Zeeman Hamiltonian and present a $g$-tensor model capturing these effects. We then use this information to show, via simulation, how control over the magnetic field angle and excitation polarization can be used to improve the amount of spin-photon entanglement generated via the protocol described in \cite{Lindner2009}.\\
\begin{figure*}[t]
    \includegraphics[width=\linewidth]{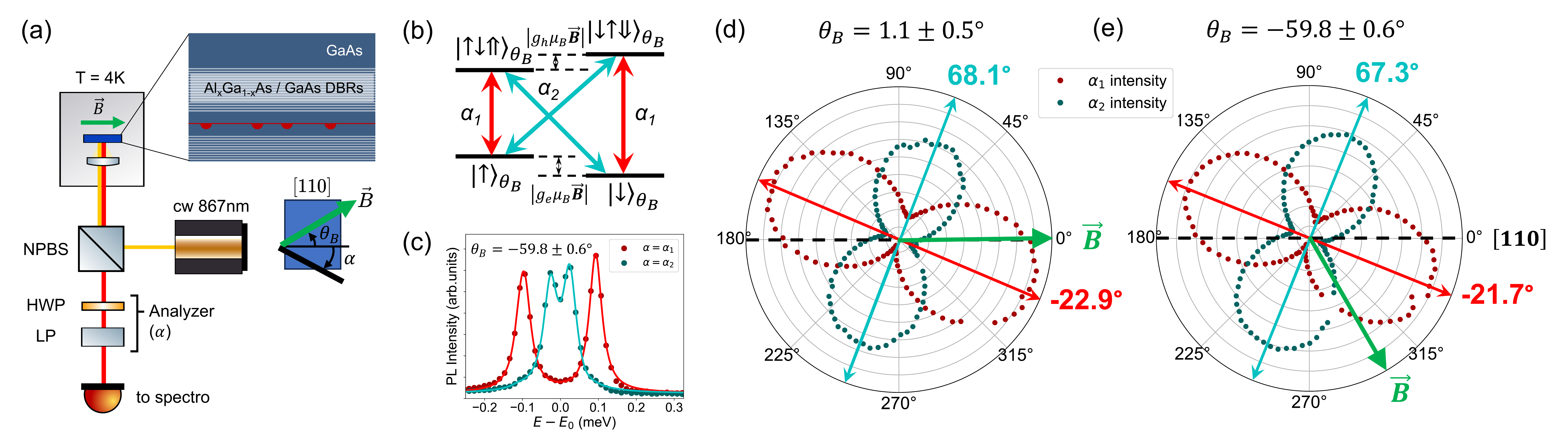}
    \caption{\label{fig:polar} (a) Experimental setup for measuring polarization-resolved PL of a single QD under a transverse magnetic field, where NPBS = non-polarizing beamsplitter, HWP = half wave plate, LP = linear polarizer. Inset shows the sample orientation with respect to the [110] crystalline axis, magnetic field angle $\theta_B$, and polarization angle $\alpha$ measured by the analyzer. (b) Energy level diagram and optical selection rules for a negative trion under a strong transverse magnetic field. (c) PL spectra for a specific value of $\theta_B$, with Lorentzian peak fitting, showing $\alpha_1$ and $\alpha_2$ eigenpolarizations. (d-e) Polar plots of PL intensity, here shown for two different angles $\theta_B$. The data points are the average integrated area of the two $\alpha_1$ peaks (dark red) or $\alpha_2$ peaks (dark teal), such as those shown in (c). Red and teal lines show the angular position of the $\alpha_1$ and $\alpha_2$ eigenpolarizations. Complete data with error is shown in Table~\ref{datatable}. The green arrow marks the real-space position of $\overrightarrow{B}$, and the black dotted line is the real-space position of the sample [110] axis (\qty{0}{\degree}).}
\end{figure*}
\indent We perform this study on a sample of InGaAs QDs embedded in a GaAs matrix, grown by molecular beam epitaxy and annealed post-growth at $\sim$\,\qty{900}{\celsius} for $\sim$\,\qty{30}{\second}, which \revisePRL{reduces the inhomogeneous broadening} and induces a blue-shift of the average PL energy via incorporation of Ga in the QD matrix~\cite{Sinha2019}. The resulting InGaAs QDs are both embedded in a n-i-p diode structure and located within a $3\lambda$ GaAs cavity between pairs of $\textrm{Al}_x\textrm{Ga}_{1-x}\textrm{As} / \textrm{GaAs}$ distributed Bragg reflectors with a cavity quality factor of approximately 8000. \\
\indent
\revisePRL{The QD studied in this work was selected at random from a representative group of bright QDs (see Appendix~\ref{sec:insitu})}. With no voltage applied\revisePRL{, the chosen QD} is host to an electron in the ground state $\ket{e^-}$, and a negative trion $\ket{X^-}$ in the excited state. We address this QD with non-resonant, above-band excitation at \qty{867}{\nano\metre} and observe PL from the QD at \qty{923.9}{\nano\metre}\revisePRL{, which is near the mean photon \revisePRL{wavelength} (\qty{925.2}{\nano\metre}) expected from an InGaAs QD annealed at \qty{900}{\celsius}~\cite{Petrov2008,Sinha2019}.
}\\
\indent We apply a transverse magnetic field of magnitude $B$ up to 5~T oriented at an angle $\theta_B$ from the crystallographic axis $[110]$ in the sample plane $(001)$  perpendicular to the optical axis $z$ (see Fig.~\ref{fig:polar}a). In practice, the sample is rotated, before cooling down to \qty{4}{\kelvin}, by the angle $-\theta_B$ with respect to the fixed horizontal magnetic field delivered by a single superconducting split-coil magnet in our optical cryostat. To measure the same individual QD at various angles $\theta_B$, we marked the position of a bright QD hosting a negative trion by drawing an oriented pattern in photoresist using an in-situ lithography technique~\cite{Dousse2008}.\\
\indent In a transverse magnetic field of this magnitude, the Zeeman splitting of the ground and excited state levels is spectrally resolved and determined by the product $|g_{e(h)}\mu_\text{B} B|$, where $\mu_\text{B}$ is the Bohr magneton and $|g_{e(h)}|$ is the effective electron (hole) $g$-factor in the ground (excited) state. This gives rise to four PL transitions of similar intensity that have orthogonal linear polarizations of angles $\alpha_1$ and $\alpha_2=\alpha_1+\pi/2$ as described in Fig.~\ref{fig:polar}b.\\
\indent To resolve the linear polarization of these PL transitions, we use an analyzer along a variable direction $\alpha$, consisting of a fixed linear polarizer and a $\lambda/2$ waveplate in a motorized rotating mount, together with a non-polarizing beam splitter to separate the collected signal from the excitation beam. A simplified schematic of the setup is shown in Fig.~\ref{fig:polar}a, with full details in Appendix~\ref{sec:setup}. Using this setup, we experimentally measure the Zeeman effect and linear polarization of the four PL transitions, as shown in Fig~\ref{fig:polar}c. The obtained polarization angles $\alpha=\alpha_1$ or $\alpha_2$, which in general differ  from the magnetic field direction, give access to the anisotropic $g$-factor.\\
\indent To understand the origin of hole $g$-factor anisotropy, we have to consider the different terms of a hole Zeeman Hamiltonian that capture the spin-dependent interaction with a magnetic field and the structural characteristics of the QD. The interaction of a hole with a magnetic field ($B$) in bulk ($T_d$ symmetry) is given by the Luttinger Hamiltonian~\cite{Luttinger1956,VanKesteren1990}:

\begin{equation}\label{eq:Lut}
    \hat{H}_{Z_h}(\textbf{B}) = -2\mu_B\sum_{i=x,y,z}(\kappa \hat{J}_i+q\hat{J}^3_i)B_i
\end{equation}

Here, $\hat{J}_i$ are the angular momentum operators for the $\Gamma_8$ valence band (including both the heavy holes $\pm3/2$ and light holes $\pm1/2$) with $i=x,y,z$, the crystallographic axes of type $\langle 100\rangle$, and $\kappa$ and $q$ the Luttinger parameters capturing their relative contributions. For a QD with in-plane dimensions $L_x\approx L_y$ significantly larger than the dimension $L_z$ along the growth direction, there is a finite splitting $\Delta_{HL}$ between light hole (LH) and heavy hole (HH) levels of the order $\sim50$~meV.\\
\indent The second contribution to the hole Zeeman splitting in a transverse field comes from the $C_{2v}$-like perturbation mentioned earlier
~\cite{Leger2007,Kowalik2007}, which gives rise to a specific mixing of HH and LH states as captured by the following Hamiltonian~\cite{Ivchenko1996,Toropov00}:
\begin{equation}
    \label{eq:c2v}
    \hat{H}_{C_{2v}} = \beta e^{-i\theta_0\hat{J}_z}\left(\hat{J}_x \hat{J}_y+\hat{J}_y \hat{J}_x\right)e^{i\theta_0\hat{J}_z}
\end{equation}
where $\theta_0$ represents the angle between one of the $C_{2v}$ mirror planes and the [110] crystalline axis. These mirror planes arise from the strain and shape anisotropy intrinsic to the QD. Eq.~\ref{eq:c2v} directly couples the $\pm3/2$ and $\mp1/2$ levels, with a magnitude captured by a parameter $\beta$. In the following, we refer to this contribution as the valence band mixing (VBM) effect. We note that analogous mixing of light- and heavy-hole states due to symmetry breaking in vertically-stacked InAs QDs leads to a variety of interesting physical phenomena with applications in scalable quantum information processing \cite{Doty2010a, Economou2012, Vezvaee2022a}. 

We can now consider the combined hole Hamiltonian of Eqs.~\ref{eq:Lut} \&~\ref{eq:c2v}, including the LH-HH splitting $\Delta_{HL}$ (full details in Appendix~\ref{sec:spinhamiltonian}). In the HH $\pm1/2$ pseudo-spin basis where $\ket{\Uparrow}\equiv\ket{S_z=+1/2}\longleftrightarrow\ket{J_z=+3/2}$ and $\ket{\Downarrow}\equiv\ket{S_z=-1/2}\longleftrightarrow\ket{J_z=-3/2}$,  the $2\times2$ matrix representing the in-plane $g$-tensor in the $x,y$ coordinate frame takes the form (to first order in $\beta/\Delta_\text{HL}$):

\begin{eqnarray}
\label{eq:gtensor}
g_h &=-3\begin{pmatrix}
q +\rho \sin2\theta_0 & \rho \cos2\theta_0 \\
-\rho \cos2\theta_0 & -q+\rho \sin2\theta_0 \\
\end{pmatrix}\\
\text{with}
    \label{eq:rho}
    &\rho = \displaystyle\frac{(4\kappa+7q)\beta}{\Delta_{\text{HL}}}
\end{eqnarray}

This expression shows that the symmetry properties of the $g$-factor are determined by the ratio $\rho /q$, which offers a way to assess the dominant contribution to the hole Zeeman splitting. We can experimentally determine this by measuring the position of the polarization eigenaxes $\alpha_1$ and $\alpha_2$ in relation to the angle $\theta_B$ of the magnetic field. Specifically, if the $C_{2v}$-like VBM effect is dominant ($\rho \gg q$) we expect the eigenpolarizations $\alpha_1$ and $\alpha_2$ of the trion transitions under a magnetic field to be fixed by the $C_{2v}$ mirror planes that are offset from the [110] crystal axis~\cite{Semenov2003}. In this case, the polarization eigenaxes remain constant in the sample reference frame with

\begin{equation}\label{eq:vbm}
    \alpha_1 \approx \theta_0
\end{equation}

In contrast, if the cubic term contribution is dominant ($q \gg \rho$), the polarization eigenaxes are determined by an interplay between the symmetry axes of the cubic crystal and the direction of the magnetic field, and thus rotate with the magnetic field but in the opposite direction~\cite{Trifonov2021a}:

\begin{equation}\label{eq:cubic}
    \alpha_1 \approx -\theta_B
\end{equation}

To experimentally determine the relationship between $\alpha_1$ and $\theta_B$, we use the polarization-resolved PL measurement described earlier. As shown in Fig~\ref{fig:polar}c, there exists a given polarization angle $\alpha_1$ ($\alpha_2$) where only the two outer (inner) PL transitions are resolved. Using the polarization analyzer shown in Fig.~\ref{fig:polar}a, we precisely identify these angles $\alpha_1$ and $\alpha_2$. This is best represented by the polar plots in Fig.~\ref{fig:polar}d-e, where we directly visualize the relationship between the [110] axes of the sample and the eigenpolarizations $\alpha_1$ and $\alpha_2$. At the small angle $\theta_B=1.1\degree$ such that the field is almost parallel to the [110] axis, we already observe a significant offset $\alpha_1=-21.7\degree$ from [110] that is incompatible with a dominant cubic term. Repeating this measurement for six different angles $\theta_B$ of the magnetic field, we observe how the detected angles of the eigenpolarizations ($\alpha_1$ and $\alpha_2$) rotate with respect to $\theta_B$. 
The results are summarized in Fig~\ref{fig:gfactors}a with full results in Appendix~\ref{sec:allpolar}. We observe that $\alpha_1$ retains a fixed offset $\theta_0$ (average $\simeq -28\degree$) relative to [110], as defined by Eq.~\ref{eq:vbm}, thus indicating that the VBM effect is dominant. From the measured eigenpolarizations $\alpha_1$ and $\alpha_2$ and the splitting of the spectrally resolved transitions, we can now derive the parameters of the hole $g$-tensor.\\
\begin{figure}
\includegraphics[width=\linewidth]{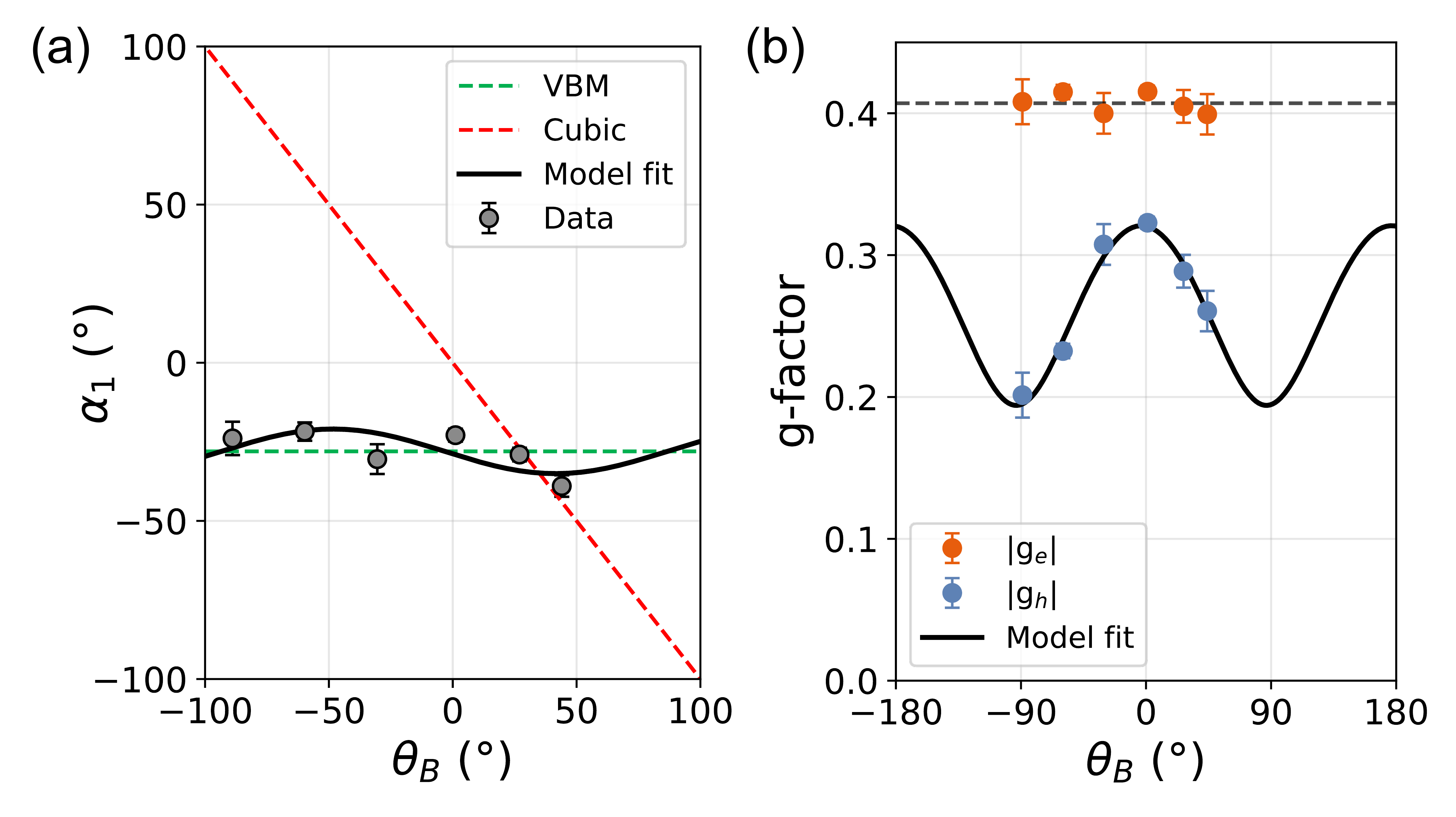}
\caption{\label{fig:gfactors}(a) Measured angle of eigenpolarization $\alpha_1$ (symbols) and $g$-factor model fit (solid line) versus the magnetic field angle $\theta_B$. Dashed green and red lines respectively show predicted relationship under dominant $C_{2v}$-driven valence-band mixing (VBM) from Eq.~\ref{eq:vbm} or dominant Luttinger $q$ parameter (cubic) from Eq.~\ref{eq:cubic}. (b) Measured $g$-factor magnitudes of the hole and electron as a function of  $\theta_B$ (symbols) along with a fit of our $g$-tensor model (solid line).}
\end{figure}
\indent We first consider the magnitudes of the electron and hole $g$-factors at each angle $\theta_B$ using the experimentally measured Zeeman splitting, \revisePRL{such as that shown in} Fig.~\ref{fig:polar}c. As \revisePRL{evidenced by} Fig~\ref{fig:gfactors}b,  we see a nearly constant value of $\abs{g_e}$ but periodic modulation in $\abs{g_h}$ as a function of $\theta_B$, which is consistent with prior results~\cite{Kaji2017,Trifonov2021a}. However, the extrema of $\abs{g_h}$ are not observed at the angles $\theta_B=-\theta_0 \mod{\pi/2}$ as determined by the tensor Eq.~\ref{eq:rho}. This discrepancy can be resolved by adding a symmetric term $\tilde{g}_h=g_h+3q_c \sigma_x$ so that the corrected tensor $\tilde{g}_h$  is no longer anti-symmetric, as expected when the QD symmetry is lower than $C_{2v}$~\cite{Serov2024} (see details in Appendix~\ref{sec:spinhamiltonian}). We note that such a term  would naturally appear by assuming a  rotation of the actual  $x$, $y$ symmetry axes of the cubic term in Eq.~\ref{eq:Lut}.  
 The dependence of $\abs{\tilde{g}_h}$ on the magnetic field angle $\theta_B$ then reads:
\begin{equation}\label{eq:geff}
    \abs{\tilde{g}_h} = 3\sqrt{\tilde{q}^2+\rho^2+2\tilde{q}\rho\cos{[2(\theta_0+\theta_c +\theta_B)]}}
\end{equation}
\noindent where $2\theta_c = \arctan(q_c/q)$ and $\tilde{q} = \sqrt{q^2 + q_c^2}$.
We then obtain a good agreement with both the $\alpha_1$ data in Fig~\ref{fig:gfactors}a and the $\abs{g_h}$ data in Fig~\ref{fig:gfactors}b, by fitting our model parameters with $q=0.010$, $\rho=0.086$, $\theta_0=-28\degree$, $q_c=0.018$. Modulo a shift in the angle dependence of $\abs{\tilde{g}_h}$, the addition of the corrective term $q_c$ still allows us to identify VBM as the dominant contribution, where the parameter $\rho$ must now be compared to $\tilde{q} =  0.021$. 
\revisePRL{This result provides important information. Specifically, prior studies of QD ensembles suggest that annealing QDs may lead to more isotropic hole $g$-factors dominated by the cubic Luttinger $q$ term~\cite{Trifonov2021a}. Our experimental results on an individual annealed QD indicate that the hole $g$-factor remains dominated by valence band mixing driven by $C_{2v}$-or-lower symmetry. This symmetry reduction and the resulting anisotropic $g$-factor are described by our $g$-tensor model for a representative annealed InGaAs QD.}\\
\begin{figure*}[t]
\includegraphics[width=\linewidth]{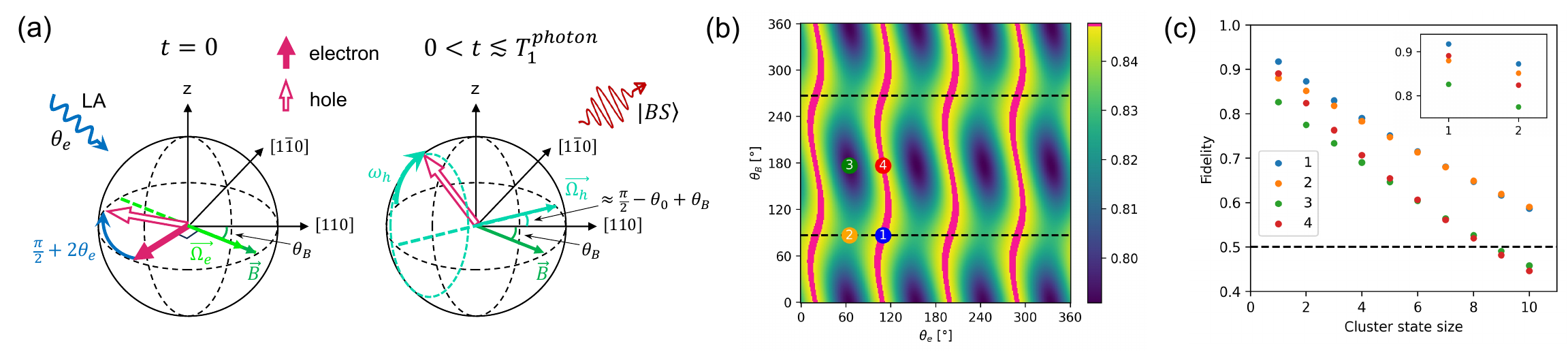}
\caption{\label{fig:fid} (a) Schematic of trion state evolution. At time $t = 0$, an excitation pulse with polarization $\theta_e$ transfers the population from the electron (ground) state to the hole (trion) state, also rotating the state about the $z$-axis. During the trion lifetime $\simeq T_1^{photon}$, the hole spin precesses at a frequency $\omega_h$ about its precession axis $\Omega_h$. (b) Concurrence plot of the spin-photon state produced by the entanglement protocol that targets $\ket{BS}$ as a function of  excitation polarization angle $\theta_e$, and magnetic field angle $\theta_B$, using the $g$-factor fit of Fig~\ref{fig:gfactors}. The red regions represent maximum entanglement, achieved when $\theta_e$ enables optimal addressing of the hole eigenstate. The dashed lines indicate the configurations where $\abs{g_h}$ (and accordingly hole state precession $\omega_h$) is minimized. Four points are marked in the plot corresponding to different limit cases of $\theta_e$ and $\theta_B$ (see main text for descriptions). (c) Scaling of cluster state fidelity for the four limit cases from (b), labelled correspondingly. Inset is focused on the top left of the plot, showing that cases \#1 and \#4 are initially highest due to optimal addressing of the correct hole eigenstate. Both (b) and (c) are directly simulated by the method described in Appendix~\ref{sec:simulation}.}
\end{figure*}
\indent \revisePRL{Using this $g$-tensor model, we probed the quality of spin-photon entanglement with such QDs} by simulating the cluster state generation protocol proposed by Ref.~\cite{Lindner2009} and implemented in Ref.~\cite{Coste2023}. The success of the entanglement is impacted by the hole $g$-factor anisotropy in two ways: 1) the precession axis of the hole spin $\vec{\Omega}_h =\tilde{g}_h \vec{B}$ might differ from $\vec{\Omega}_e =g_e \vec{B}$ of the electron spin, and 2) the magnitude of $|\tilde{g}_h|$ controls the magnitude of the trion hole spin precession (and thus the relative dephasing between the entangled spin-photon states). Both of these effects are depicted in Fig.~\ref{fig:fid}a.\\
\indent We consider the QD electron spin in a weak ($60$ mT) transverse magnetic field such that both the fast quasi-resonant excitation  and photon detection are not sensitive to the small Zeeman splitting. In the usual convention of a $1/2$ pseudo-spin for the heavy-hole, the optical selection rules in the $z$ basis are 
$\ket{\uparrow}\,\longleftrightarrow\,\ket{\Uparrow}$ and $\ket{\downarrow}\,\longleftrightarrow\,\ket{\Downarrow}$, mapping respectively to circular right $\left(\ket{R}\right)$ and circular left $\left(\ket{L}\right)$ photon polarizations.
The QD is assumed to be addressed quasi-resonantly with polarization-preserving optical pulses, such as with the longtitudinal-acoustic phonon (LA) excitation scheme~\cite{Thomas2021,Coste2023a}.\\
\indent In order to perform the first entangled emission cycle of the cluster state procedure, the electron spin is heralded in the pure state $\ket{\uparrow}$ when a first photon is detected in the polarization state $\ket{R}$ at $t=- \pi/2\omega_e$. A second pulse is then synchronized to excite the QD at $t=0$ after a $\pi/2$ precession of the spin about the in-plane magnetic field $\vec{B}$ applied at an angle $\theta_B$ with respect to the direction $[110]$. In other words, the second excitation pulse arrives when the electron spin state is $\ket{e}=\frac{1}{\sqrt{2}}\left(\ket{\uparrow}+i e^{i(\theta_B+\pi/4)}\ket{\downarrow}\right)$.
If this second LA excitation pulse is linearly polarized with some angle $\theta_e$ with respect to $[110]$, the resulting excitation event leads to a hole spin $\left(\cos(\theta_e+\pi/4) (\hat{\sigma}^\dagger_R+\hat{\sigma}^\dagger_L)+\sin(\theta_e+\pi/4) i(\hat{\sigma}^\dagger_L-\hat{\sigma}^\dagger_R)\right)\ket{e}$ where $\hat{\sigma}^\dagger_R=\ket{\Uparrow}\bra{\uparrow}$ and $\hat{\sigma}^\dagger_L=\ket{\Downarrow}\bra{\downarrow}$ are the jump operators mapping the electron spin to the hole spin according to the optical selection rules.\\
\indent As a result, the polarized excitation transfers the electron spin state to the following pure hole (trion) spin state, as depicted in Fig.~\ref{fig:fid}a at time $t=0$:\\
\begin{equation}
\label{eq:holestate}
|h\rangle=\frac{1}{\sqrt{2}}\left(e^{-i(\theta_e+\pi/4)}\ket{\Uparrow}- e^{i(\theta_e+\theta_B)}\ket{\Downarrow}\right)
\end{equation}

In the ideal limit where $\omega_h\equiv\abs{\tilde{g}_h}\mu_\text{B}B/\hbar$ verifies $\omega_h T_1^{photon} \ll 1$, the hole state $|h\rangle$ recombines instantly, producing the spin-photon Bell state $|BS\rangle=\frac{1}{\sqrt{2}}\left(\ket{\uparrow}\ket{R}-e^{i(2\theta_e+\theta_B+\pi/4)}\ket{\downarrow}\ket{L}\right)$ and concluding the cycle. However, as illustrated in Fig.~\ref{fig:fid}a, for a finite trion lifetime $t\lesssim T_1^{photon}$ there is indeed some unwanted hole precession about its axis $\Omega_h$.\\
\indent Considering a \revisePRL{realistic} trion source with the hole $g$-factor $\tilde{g}_h$ extracted from the fit in Fig~\ref{fig:gfactors}, we explicitly simulate the procedure for entangled emission described above \revisePRL{ with a magnetic field $B= \qty{60}{\milli\tesla}$} (see Appendix~\ref{sec:simulation}), using concurrence to quantify entanglement in the resulting spin-photon state.
The behaviour of this simulated value of concurrence with respect to the choice of $\theta_B$ and $\theta_e$, depicted in Fig.~\ref{fig:fid}b, highlights the impact of the $g$-factor anisotropy on the quality of spin-photon entanglement.
Considering equation \ref{eq:holestate}, control over the protocol parameters $\theta_e$ and $\theta_B$ modifies the complex phase between the two components of the excited hole state, i.e.~rotates the spin state in the equatorial plane of its Bloch sphere. Notably, for a given magnetic field angle $\theta_B$, the polarization angle $\theta_e$ can always be chosen such that  the hole spin $|h\rangle$ is an eigenstate of the hole Zeeman Hamiltonian and therefore does not precess during the photon emission.
The configurations of $\theta_e$ targeting a Zeeman eigenstate minimize hole precession-induced dephasing and result in maximum concurrence, depicted in the red regions in Fig.~\ref{fig:fid}b. This maximum concurrence is then only limited by the electron spin precession during the radiative lifetime. The dependence of $|\tilde{g}_h|$ on the angle of the magnetic field (see Eq.~\ref{eq:geff}) is instead responsible for the modulation on the other axis: the magnitude of $|\tilde{g}_h|$ controls the magnitude of hole precession and thus dictates only how poor the concurrence is whenever $\theta_e$ is not optimized to target a Zeeman eigenstate. We observe two values of $\theta_B$ where concurrence is globally higher, namely $\theta_B \approx \pm90\degree$, which correspond to the two angles where $|\tilde{g}_h|$ is at its minimum. This effect of anisotropy reaches its peak when $\rho=\tilde{q}$, such that the form of the $g$-tensor allows two configurations where $|\tilde{g}_h| = 0$ (depicted in Fig.~\ref{fig:landscape}).\\
\indent Looking beyond spin-photon Bell states, we simulate how the hole $g$-factor affects the generation of larger cluster states (i.e.~spin-multiphoton entangled states). We identify in Fig.~\ref{fig:fid}b four distinct regimes to operate the entanglement cycle: according to the choice of $\theta_B$ and $\theta_e$ the cycle can take place in a condition of maximum (2,3) or no hole precession (1,4), owing to correct or incorrect addressing of the hole eigenstate, and maximum (3,4) or minimum (1,2) coupling to the magnetic field, from tuning of $\abs{\tilde{g}_h}$ via $\theta_B$.\\
\indent As we are considering multipartite states, we therefore use the fidelity of the generation procedure as an indirect quantifier of entanglement in the final state. In particular we probe how the choice of regime impacts its scaling with the size of the cluster state. The results of Fig.~\ref{fig:fid}c show that tuning $\theta_e$ to reach the hole eigenstate condition enables the higher initial fidelity in regimes 1 and 4 (see inset). However, the minimization of $|\tilde{g}_h|$ in regimes 1 and 2 significantly impacts the fidelity as the cluster state size increases. While the improvement provided by optimal $\theta_e$ eventually diminishes, a substantial difference remains between optimal (regimes 1 and 2) and suboptimal (regimes 3 and 4) magnetic field angles. Optimal scaling up to larger cluster states will thus require tuning of both the magnetic field angle and the polarization of excitation.\\
\indent \revisePRL{Even though only one QD in the $X^-$ configuration was examined in detail, our conclusions on the role of the hole $g$-factor in cluster state generation also apply to the case of a positive trion ($X^+$), which is more desirable due to the longer spin coherence time of the hole. In this case we would wish to minimize the spin precession frequency in the excited state $(\omega_e\propto B)$ with respect to the radiative lifetime to yield again an improvement of cluster state fidelity similar to that calculated in Fig.~\ref{fig:fid}c. Considering a pulse sequence synchronized to the hole spin precession $(\omega_h\propto |g_h|B)$ in the ground state, the required magnetic field $B$ (and therefore $\omega_e$) could be minimized by now maximizing $|g_h|$ using the magnetic field angle, as determined by our $g$-tensor model.} Further studies could also introduce electric fields or strain to tune the $g$-factors~\cite{Doty2006, Pingenot2008,Bennett2013,Liu2011,Prechtel2015,Tholen2016}. In-situ rotation of the angle between the sample and magnetic field using either a vector magnet or a piezo rotator would have allowed exploration of more angles with more precise control, and indeed such control will likely be needed in the future to maximize fidelity of cluster states. Finally, while inferred measurements of entanglement concurrence using continuous wave excitation have been conducted recently~\cite{Serov2024}, we could explicitly validate our simulations by generating and measuring multipartite cluster states under various excitation polarizations and magnetic field angles.\\
\indent In summary, we investigated the in-plane hole $g$-factor \revisePRL{of an} annealed InGaAs QD and the practical implication of its anisotropy for spin-photon entanglement. We showed experimentally that the polarization eigenaxes of the Zeeman-split transitions remain mostly locked with respect to the sample $[110]$ axes by a fixed offset $\theta_0$, irrespective of the angle of the applied magnetic field $\theta_B$. This tells us that the valence band mixing effect is the dominant contribution to the hole $g$-tensor anisotropy \revisePRL{in spite of annealing}. The hole $g$-factor tensor extracted from this experimental data allowed us to simulate spin-photon entanglement. Using this simulation, we showed how the choice of the magnetic field angle $\theta_B$ and the polarization of excitation $\theta_e$ affect the quality of entanglement, ultimately by controlling the position and precession frequency of the \revisePRL{trion spin in its Bloch sphere}. We found that maximal concurrence is always achievable in a single entangled emission step through the addressing of the \revisePRL{trion spin} eigenstate by tuning only $\theta_e$, but fidelity of multipartite cluster states at optimal $\theta_B$ scales almost identically irrespective of $\theta_e$, due to the \revisePRL{trion spin} precession being minimized to globally achieve higher entanglement. These results elucidate the specific impact of hole $g$-factor anisotropy on spin-photon entanglement, and suggest how this anisotropy can be harnessed to optimize the generation of larger cluster states.

\begin{acknowledgments}
The authors wish to acknowledge H\^elio Huet, Nathan Coste, and Manuel Gundin for their useful inputs. This work was partially supported by the Paris Ile-de-France Région in the framework of DIM SIRTEQ, the Research and Innovation Programme QUDOT-TECH under the Marie Skłodowska-Curie grant agreement 861097, the French National Research Agency  (ANR) project ANR-21-CE30-0049-02, the CEFIPRA  project 64T3-2, the Plan France 2030 through the projects ANR22-PETQ-0011 and ANR-22-PETQ-0013, and a public grant overseen by the French National Research Agency (ANR) as part of the ”Investissements d’Avenir” programme (Labex NanoSaclay, reference: ANR\-10\-LABX\-0035). \revisePRL{This work has also been co-funded by the Horizon-CL4 program under the grant agreement 101135288 for the EPIQUE project.} P.R.R. acknowledges the financial support of the Fulbright-Universit\'{e} Paris-Saclay Doctoral Research Award and the UNIDEL Distinguished Graduate Fellowship. M.F.D. acknowledges support from the National Science Foundation (2217786). This work was done within the C2N micro nanotechnologies platforms and partly supported by the RENATECH network and the General Council of Essonne.

\end{acknowledgments}

\bibliography{library}


\appendix

\clearpage

\begin{widetext}

\counterwithin{figure}{section}
\counterwithin{table}{section}
\renewcommand\thefigure{\thesection\arabic{figure}}
\renewcommand\thetable{\thesection\arabic{table}}
\renewcommand\theequation{\thesection\arabic{equation}} 

\revisePRL{\section{\label{sec:insitu}Identification of individual quantum dots}

An in-situ lithography technique~\cite{Dousse2008} was used to mark the position of quantum dots for this study. Photoluminescence spectroscopy was first utilized to map QDs emitting in the 920-930 nm range, of which there are millions in a given sample area of \qty{0.12}{\centi\metre\squared}. Those that had bright PL emission with zero applied electrical bias were chosen for this study for ease of experimentation, to avoid sample rotation with electrical wiring. \revisePRL{For this purpose, 16 QDs were marked in a chosen $\sim\qty{1}{\milli\metre\squared}$ region of the sample, a region normally containing on the order of thousands of dots.} A microscope image of the marked QDs and their corresponding PL spectra are shown in Figure~\ref{fig:insitu}.

\begin{figure*}[h]
\includegraphics[scale=0.13]{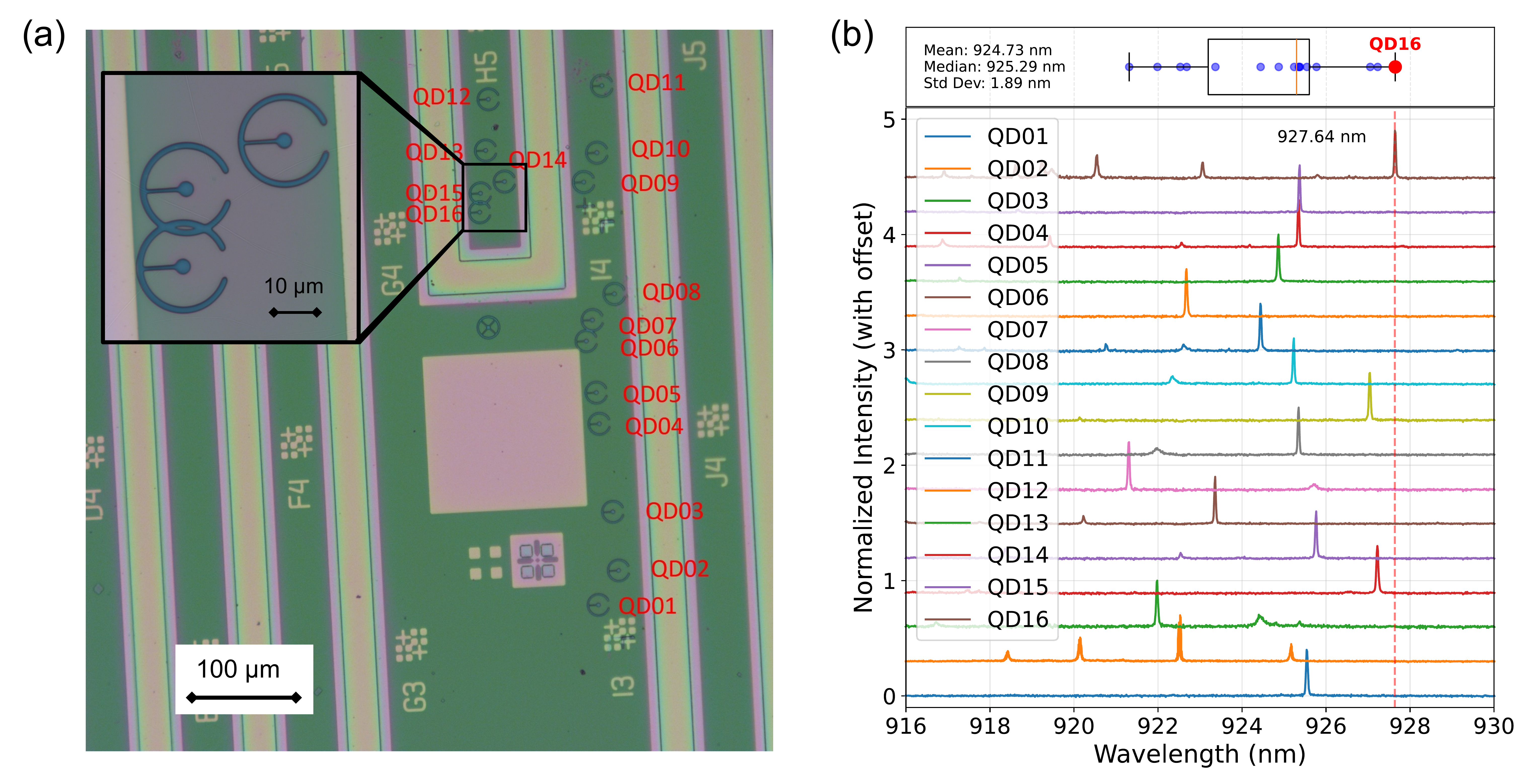}
\caption{\label{fig:insitu}(a) Optical microscope image of sixteen QDs with lithographic marks. Each QD is located at the center point of its respective mark. (b) PL spectra of the sixteen QDs taken prior to lithographic development, offset for clarity. Above: Corresponding box plot of the peak emission wavelengths of all of the QDs.}
\end{figure*}

The QD used in this study is QD16 with a PL peak emission wavelength of \qty{927.64}{\nano\metre}. This was selected from the sample set at random. The statistical analysis of the emission wavelengths of the selected sample of QDs, plotted in Figure~\ref{fig:insitu}b, shows a median in line with expectation for an InGaAs QD annealed at \qty{900}{\celsius}~\cite{Sinha2019,Petrov2008} and does not present any outliers, suggesting that the group from which we extracted QD16 is representative. We note that the peak wavelength of QD16 as reported in the main manuscript is \qty{923.9}{\nano\metre}. This difference is due to a combination of the lithographic processing and modest differences in the sample temperature between the two optical cryostats where the in-situ lithography and magneto-optical PL experiments were conducted.}

\section{\label{sec:allpolar}Complete measurements of polarization eigenaxes}

Below are the complete polar plots for each of the six magnetic field angles ($\theta_B$) measured. The values of $\alpha_1$ from these plots are shown in Fig.~\ref{fig:gfactors}a. These plots show the PL intensity of each of the four PL transitions, noting that $\alpha_1$ ($\alpha_2$) and $\alpha'_1$ ($\alpha'_2$) have the same linear polarization but different energies. These differ from the plots in Fig.~\ref{fig:polar}d and e, where the average intensity of the two transitions with the same polarization (e.g. $\alpha_1$ and $\alpha'_1$) is instead plotted for clarity.

\begin{figure*}[h]
\includegraphics[scale=0.136]{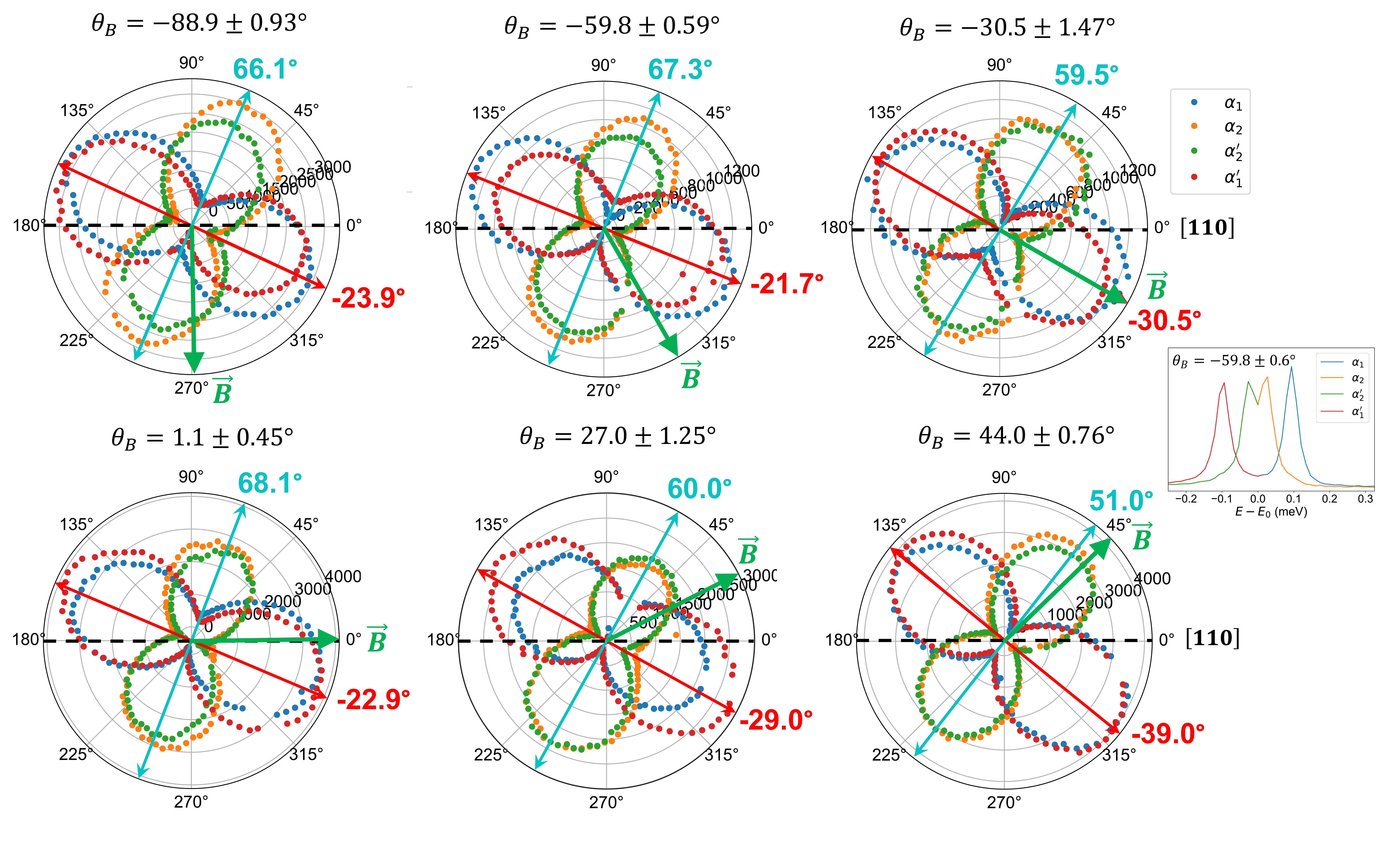}
\caption{\label{fig:allpolar}Polar plots of the intensity of the four Zeeman split transitions for six different angles $\theta_B$, marked above. Inset shows PL spectrum for a given $\theta_B$ with each of the four transitions marked with their respective polarization. Marked values of $\alpha_1$ and $\alpha_2$ do not include error bars. Complete data with error is shown in Table~\ref{datatable}.}
\end{figure*}

The complete data set of key parameters $\alpha_1$, $\alpha_2$, $\abs{g_h}$ and $\abs{g_e}$ for each angle $\theta_B$ is summarized below, with error bars. 

\begin{table}[h!]
    \label{datatable}
    \centering
    \renewcommand{\arraystretch}{1}
    \setlength{\arrayrulewidth}{0.5mm}
    \arrayrulecolor{black}
    \setlength{\tabcolsep}{10pt}
    \begin{tabular}{|c | c | c | c | c | c |}
        \hline
        $\theta_B$ (\degree) & $\alpha_1$ (\degree) & $\alpha_2$ (\degree) & $|g_e|$ & $|g_h|$ \\ \hline
        $1.1 \pm 0.45$ & $-22.9 \pm 2$ & $68.10 \pm 3.3$ & $0.415 \pm 0.004$ & $0.323 \pm 0.004$ \\ \hline
        $-30.5 \pm 1.5$ & $-30.5 \pm 4.7$ & $59.5 \pm 4.3$ & $0.400 \pm 0.014$ & $0.308 \pm 0.014$ \\ \hline
        $27 \pm 1.3$ & $-29 \pm 2.1$ & $60 \pm 2.1$ & $0.405 \pm 0.012$ & $0.289\pm 0.012$ \\ \hline
        $-59.8 \pm 0.59$ & $-21.8 \pm 2.9$ & $67.3 \pm 3.4$ & $0.415 \pm 0.005$ & $0.232 \pm 0.005$ \\ \hline
        $44 \pm 0.76$ & $-39 \pm 3.4$ & $51.0 \pm 3.4$ & $0.399 \pm 0.014$ & $0.261 \pm 0.014$ \\ \hline
        $-88.9 \pm 0.93$ & $-23.9 \pm 5.3$ & $66.1 \pm 5.3$ & $0.408 \pm 0.017$ & $0.201 \pm 0.016$ \\ \hline
    \end{tabular}
\end{table}

\section{\label{sec:spinhamiltonian} Spin Hamiltonian model}

Our spin model considers an isotropic electron $g$-factor and anisotropic hole $g$-factor. The electron Zeeman Hamiltonian is defined as 
\begin{equation}
    \hat{H}_{Z_e}(\textbf{B}) = g_e\mu_B\left(S_x B\cos(\theta_B+\pi/4)+S_y B\sin(\theta_B+\pi/4)+S_ZB_z\right)
\end{equation}
where $S_x,S_y,S_z$ are the $\frac{1}{2}$ spin Pauli matrices as follows. 

\[
S_x = \frac{1}{2} \begin{pmatrix} 0 & 1 \\ 1 & 0 \end{pmatrix}, \quad
S_y = \frac{1}{2i} \begin{pmatrix} 0 & 1 \\ -1 & 0 \end{pmatrix}, \quad
S_z = \frac{1}{2} \begin{pmatrix} 1 & 0 \\ 0 & -1 \end{pmatrix}
\]

Our model of the hole spin begins by examining the Hole Zeeman effect. In Fig~\ref{fig:holezeeman}, we represent the linear Zeeman splitting $\Delta_{Z_h}$ between the heavy hole levels $J_z=\pm\frac{3}{2}$. However to understand this splitting we must consider all four hole energy levels, heavy hole $Jz=\pm\frac{3}{2}$ and light hole $J_z=\pm\frac{1}{2}$, which are split by $\Delta_{HL}\approx50\textrm{meV}$ due to the strong confinement of the carriers along the QD growth axis $z$. To the zeroth order, the effect of spatial confinement can be represented by a  $D_{2d}$ Hamiltonian which is still diagonal in the basis of  $\hat{J}_z$ eigenstates and simply reads:
\begin{equation}
    \hat{H}_{D_{2d}} = -\frac{\Delta_{HL}}{3}\left(J^2_z - \frac{\hat{J}_x\hat{J}_y+\hat{J}_y\hat{J}_x}{2}\right)
\end{equation}

\begin{figure}[h]
\includegraphics[scale=0.12]{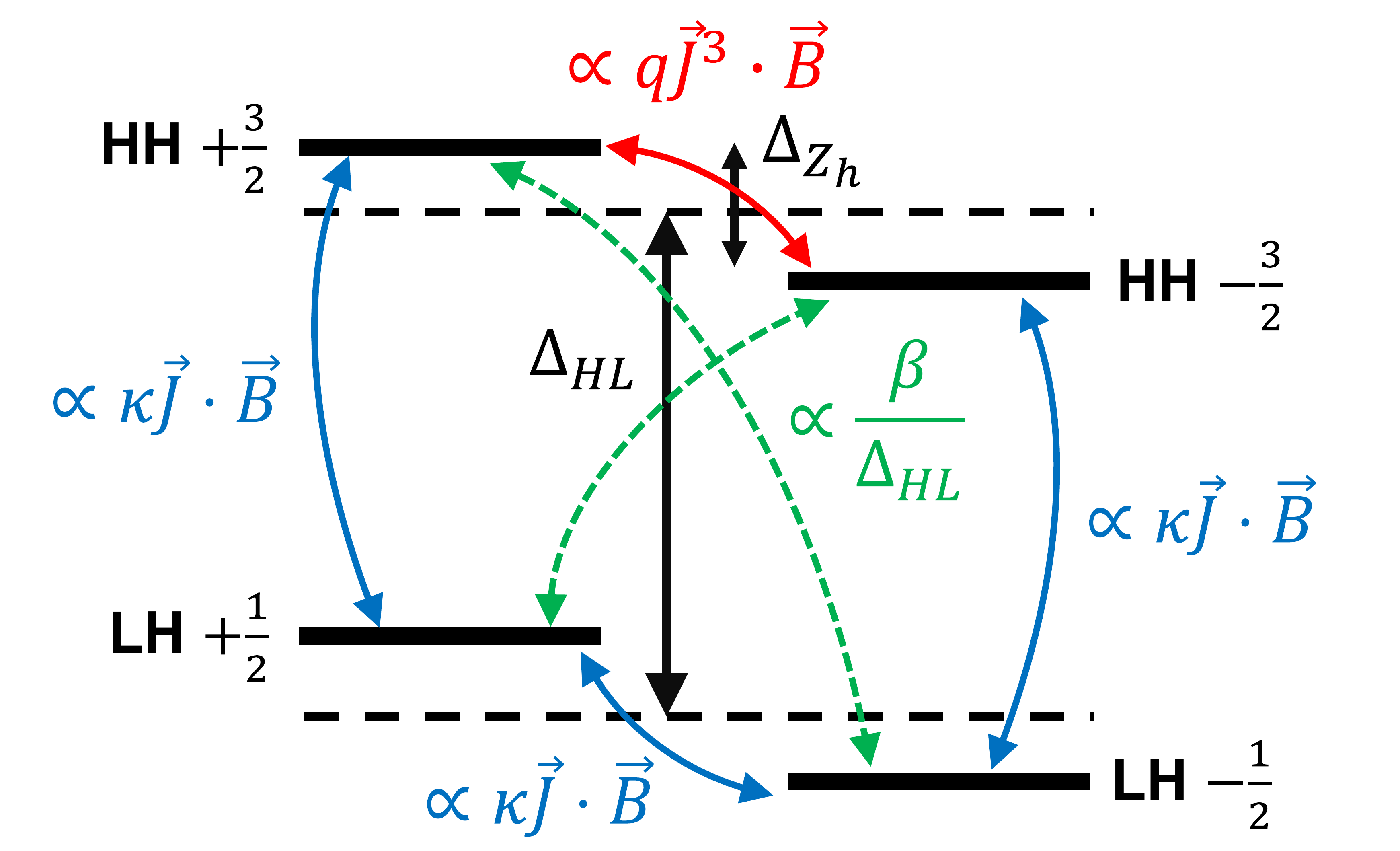}
\caption{\label{fig:holezeeman}Schematic of the hole Zeeman energy levels depicting the splitting $\Delta_\text{HL}$ between the QD ground state heavy hole and the first QD  light-hole state along with  the coupling between these  levels due to the magnetic field or symmetry reduction. Blue arrows refer to the coupling from the linear term ($\kappa$) of the Luttinger Hamiltonian, red refers to the cubic term ($q$), and green refer to the coupling arising from $C_{2v}$-driven valence band mixing Hamiltonian.}
\end{figure}

First we look at the Luttinger Hamiltonian from Eq.~\ref{eq:Lut}:
\begin{equation}
    \hat{H}_{Z_h}(\textbf{B}) = -2\mu_B\sum_{i=x,y,z}(\kappa \hat{J}_i+q\hat{J}^3_i)B_i,
\end{equation}
where the operators $\hat{J}_x,\hat{J}_y,$and $\hat{J}_z$ represent the components of the angular momentum of the hole total spin $J=3/2$ along the directions $x,y,z$ associated to the crystallographic axes $\langle 100\rangle$. In the canonical basis of the hole spin, the matrices representing these operators read:

\[
J_x = \frac{1}{2} \begin{pmatrix} 0 & \sqrt{3} & 0 & 0 \\ \sqrt{3} & 0 & 2 & 0 \\ 0 & 2 & 0 & \sqrt{3} \\ 0 & 0 & \sqrt{3} & 0 \end{pmatrix}, \quad
J_y = \frac{1}{2i} \begin{pmatrix} 0 & \sqrt{3} & 0 & 0 \\ -\sqrt{3} & 0 & 2 & 0 \\ 0 & -2 & 0 & \sqrt{3} \\ 0 & 0 & -\sqrt{3} & 0 \end{pmatrix}, \quad
J_z = \begin{pmatrix} 3/2 & 0 & 0 & 0 \\ 0 & 1/2 & 0 & 0 \\ 0 & 0 & -1/2 & 0 \\ 0 & 0 & 0 & -3/2 \end{pmatrix}
\]
We can see the $\hat{J}_x$ and $\hat{J}_y$ matrices do not directly couple the heavy hole levels $+\frac{3}{2}$ and $-\frac{3}{2}$. They instead open up $\Delta_{Z_h}$ via the indirect coupling to the light-hole levels, as shown by the blue terms and arrows in Fig~\ref{fig:holezeeman}. The corresponding effect  which  scales as $\abs{B}^3/\Delta_\text{HL}^2$, is thus strongly reduced by the splitting  $\Delta_\text{HL}$ and moreover does not provide  the linear dependence in $\abs{B}$ as observed empirically. 

The second term of this Hamiltonian, containing the Luttinger $q$ parameter, captures the magnitude of coupling from the angular momentum matrices raised to the third power: 
\[
J_x^3 = \begin{pmatrix} 
0 & \frac{7\sqrt{3}}{8} & 0 & \frac{3}{4} \\ 
\frac{7\sqrt{3}}{8} & 0 & \frac{5}{2} & 0 \\ 
0 & \frac{5}{2} & 0 & \frac{7\sqrt{3}}{8} \\ 
\frac{3}{4} & 0 & \frac{7\sqrt{3}}{8} & 0 
\end{pmatrix}, \quad
J_y^3 = \begin{pmatrix} 
0 & -\frac{7i\sqrt{3}}{8} & 0 & \frac{3i}{4} \\ 
\frac{7i\sqrt{3}}{8} & 0 & -\frac{5i}{2} & 0 \\ 
0 & \frac{5i}{2} & 0 & -\frac{7i\sqrt{3}}{8} \\ 
-\frac{3i}{4} & 0 & \frac{7i\sqrt{3}}{8} & 0 
\end{pmatrix}, \quad
J_z^3 = \begin{pmatrix} 
\frac{27}{8} & 0 & 0 & 0 \\ 
0 & \frac{1}{8} & 0 & 0 \\ 
0 & 0 & -\frac{1}{8} & 0 \\ 
0 & 0 & 0 & -\frac{27}{8} 
\end{pmatrix}
\]
Here, there is a direct coupling between $J_z=\pm\frac{3}{2}$ levels that would be responsible for linear $\Delta_{Z_h}$, as indicated by the red symbols and arrows in Fig~\ref{fig:holezeeman}.

Finally, we look at the Hamiltonian capturing the $C_{2v}$-like valence band mixing effect, from Eq.~\ref{eq:c2v}: \begin{equation}
    \hat{H}_{C_{2v}} = \beta e^{-i\theta_0\hat{J}_z}\left(\hat{J}_x \hat{J}_y+\hat{J}_y \hat{J}_x\right)e^{i\theta_0\hat{J}_z}
\end{equation}
where $\beta$ is a real parameter capturing the strength of the mixing and $\theta_0$ is the  angle between one of the $C_{2v}$ mirror planes and the [110] crystalline axis.  This Hamiltonian directly couples the $\pm\frac{3}{2}$ and $\mp\frac{1}{2}$ levels, as shown in green in Fig.~\ref{fig:holezeeman}, such that in combination of the Luttinger $\kappa$ term contributes also to the linear scaling of $\Delta_{Z_h}$ with the magnetic field.

To define our $g$-tensor model, we consider the combination of these three Hamiltonians as follows.
\begin{equation}
    \hat{H}_h = \hat{H}_{Z_h}(\textbf{B})+\hat{H}_{D_{2d}}+\hat{H}_{C_{2v}}
\end{equation}

Then as explained in the text, we cast HH levels as $\pm\frac{1}{2}$ pseudospins, with the convention that $\ket{\Uparrow}\equiv\ket{S_z=+1/2}\longleftrightarrow\ket{J_z=+3/2}$ and $\ket{\Downarrow}\equiv\ket{S_z=-1/2}\longleftrightarrow\ket{J_z=-3/2}$. Since the $4\times4$ hole Hamiltonian $\hat{H}_h$ cannot be diagonalized analytically, we first diagonalize the $\hat{H}_{C_{2v}}$ part and then express the Zeeman term $\hat{H}_{Z_h}$ in the  basis of $\hat{H}_{C_{2v}}$  eigenstates, which provides a direct term coupling the effective heavy-hole ground states. From this coupling, we derive the expression for the $g_h$ tensor to the first order in $\beta/\Delta_\text{HL}$ shown in Eq.~\ref{eq:gtensor}.
Adding the term $q_c$ to account for lower than $C_{2v}$ symmetry~\cite{Serov2024} results in the following corrected $g$-tensor:

\begin{equation}
\label{eq:corrgtensor}
\tilde{g}_h =-3\begin{pmatrix}
q +\rho \sin2\theta_0 & \rho \cos2\theta_0 -q_c\\
-\rho \cos2\theta_0 -q_c & -q+\rho \sin2\theta_0
\end{pmatrix}
\end{equation}
  
Such \revisePRL{a} $g$-tensor can be decomposed into the sum of two independent contributions: 
\begin{equation}\label{eq:gdec}
\tilde{g}_h = -3
\begin{pmatrix}
q & -q_c\\
-q_c & -q \\
\end{pmatrix}
- 3\rho
\begin{pmatrix}
 \sin2\theta_0 & \cos2\theta_0 \\
-\cos2\theta_0 & \sin2\theta_0 \\
\end{pmatrix}
\end{equation}

The contribution proportional to $\rho$ can be directly identified with a rotation matrix of an angle $2\theta_0-\pi/2$. The first contribution can instead be rewritten as:

\begin{equation}
3\tilde{q}
    \begin{pmatrix}
     \sin2\theta_c & -\cos2\theta_c \\
     \cos2\theta_c & \sin2\theta_c \\ 
    \end{pmatrix}
    \begin{pmatrix}
    0 & 1 \\
    1 & 0\\ 
    \end{pmatrix}
\end{equation}

Where $\tilde{q} = \sqrt{q^2 + q_c^2}$ and $2\theta_c = \arctan\frac{q_c}{q}$, namely a rotation by an angle $\pi/2-2\theta_c$ times an axis exchange.
Now the effective coupling strength $\abs{\tilde{g}_h}$ is computed considering the definition $\abs{\tilde{g}_h}|\vec{B}| = |\vec{\Omega}_h| = |\tilde{g}_h \cdot \vec{B}|$.
The decomposition shown above allows to interpret the quantity $\vec{\Omega}_h$ as the vector sum of two rotated versions of $\vec{B}$, proportional to $\tilde{q}$ and $\rho$.
The modulus of such vector follows from the application of cosines law, noting that for $\tilde{q}\rho \geq 0$ the orientation of the two components differs by an angle $2(\theta_0 + \theta_c + \theta_B)$:
\begin{equation}
    |\vec{\Omega}_h| = |\vec{B}| \: 3\sqrt{(\tilde{q})^2+\rho^2+2\tilde{q}\rho\cos{[2(\theta_0+\theta_c +\theta_B)]}}
\end{equation}

Division by $|B|$ directly results in Eq.~\ref{eq:geff}, reproduced below.
\begin{equation}
    \abs{\tilde{g}_h} = 3\sqrt{(\tilde{q})^2+\rho^2+2\tilde{q}\rho\cos{[2(\theta_0+\theta_c +\theta_B)]}}
\end{equation}

We note the corrective term $q_c$ adds an angle shift $\theta_c=31.3\degree$ to $\theta_0$ in the effective angle dependence of $|g_h|$. It has however no impact on the directions of polarization eigenaxes which remain mostly determined by $\alpha_1\approx\theta_0$ and $\alpha_2\approx\theta_0+\pi/2$ (in the limit $\rho\gg \tilde{q}$, we find $\alpha_1=\theta_0-\tilde{q}\sin[2(\theta_B + \theta_0+\theta_c)]/2\rho$).

\section{\label{sec:simulation} Simulation of spin-photon entanglement}

The protocol for spin-photon entanglement is nearly exactly as implemented in Ref.~\cite{Coste2023}. The system is a single QD charged with a single electron. The QD is addressed with polarization preserving, quasi-resonant excitation, such as longitudinal acoustic (LA) phonon excitation. Experimentally, this protocol would be quite simple, requiring a weak transverse magnetic field ($<100$~mT) to induce spin precession, and two linearly-polarized excitation pulses. The first pulse triggers emission of a photon, which when detected in $R$ or $L$, heralds the spin state in pure state $\ket{\uparrow} (R)$ or $\ket{\downarrow} (L)$. The spin precesses for a time equivalent to a $\pi/2$ rotation, and then a second pulse triggers emission of a photon entangled between spin and photon polarization. 

A longer sequence of such pulses results in the protocol for linear cluster state generation. In this case the heralding step is followed by $n$ $\pi/2$ rotations along with $n$ triggered photon emission cycles. The procedure is concluded by waiting for the spin to perform a $\pi$ rotation, the triggered emission of a last photon and its subsequent detection in either $L$ or $R$. This last measurement decouples the spin state from the rest of the system, resulting in the production of a purely photonic $n$-qubit linear cluster state.

Considering the anisotropic hole $g$-tensor adds some complexity to these processes. First, the magnetic field is assumed to be at some angle $\theta_B$ from the [110] crystalline axis. The electron spin therefore has a preferential axis $\Omega_e$ about which it precesses when the magnitude of the magnetic field is greater than zero. Once it completes a $\pi/2$ rotation, an excitation pulse with linear polarization at angle $\theta_e$ arrives, transferring the population to the trion (hole) state, with a rotation of $2\theta_e$ in the equatorial plane of the Bloch sphere preparing the following state:

\begin{equation}
\ket{h} = \frac{1}{\sqrt{2}}\left(e^{-i(\theta_e+\pi/4)}\ket{\Uparrow}_z- e^{i(\theta_e+\theta_B)}\ket{\Downarrow}_z\right)
\end{equation}

Given the optical selection rules this state would ideally decay into 
$
    \frac{1}{\sqrt{2}}\left(\ket{\uparrow}_z\ket{R} - e^{i(2\theta_e+\theta_B+\pi/4)}\ket{\downarrow}_z\ket{L}\right)
$,
However, realistic considerations ought to consider that both the electron and hole state evolve during the non-zero radiative lifetime $T_1^{photon}$.
This process ultimately entangles the excited state population, final spin state, and the spin-photon phase $e^{i(2\theta_e+\theta_B+\pi/4)}$ to the exact instant at which the photon was emitted.
If the time-of-arrival of such photon is not resolved on detection, as in our setup, this form of entanglement degrades into decoherence with a magnitude proportional to the product between the electron (hole) precession rate $\omega_e$ $(\omega_h)$ and the radiative lifetime $T_1^{photon}$.

To assess these effects we explicitly simulate the trion system. We employed the same master equation solving techniques detailed in the Supplementary Information of Ref. \cite{Coste2023}, which leverage a time integrated method based on simulating the dynamics of the light-matter system conditioned to not emitting any photons \cite{Wein2024}.
Akin to what one would measure experimentally with a tomography, this method allows the efficient computation of the reduced (i.e. time integrated) density matrix of the spin-photon state.

The base Hamiltonian considered for the solid-state part of the master equation reads:
\begin{equation*}
    H = \frac{\mu_B}{2} \sum_{i \in \{x,y,z\}}S_i \cdot g_e \cdot B^{(e)}_i + \frac{\mu_B}{2} \sum_{i,j \in \{x,y,z\}} S_i \cdot \left(g_h\right)_{ij} \cdot B^{(h)}_j
\end{equation*}

\noindent where $S^{(\hdots)}_i$ are the Pauli operators of the (e)lectron ($\ket{\uparrow}, \ket{\downarrow}$) and (h)ole spin state ($\ket{\Uparrow}, \ket{\Downarrow}$), $g_e, g_h$  are the tensors specifying their coupling with the external magnetic field vector $\vec{B}$.
As described in the main text $g_e$ is considered as isotropic while the form of $g_h$ is specified in terms of the parameters $q$, $\rho$, $\theta_0$ and $q_c$ (Eq.~\ref{eq:corrgtensor}).
The impact of spontaneous emission in the two light-modes is then captured with the two Lindblad dissipator terms $\mathcal{D}_{\sigma_L}, \mathcal{D}_{\sigma_R}$ in the full master equation: 
\begin{equation*}
    \frac{d}{dt}\rho(t) = -\frac{i}{\hbar}\left[H,\rho(t)\right]+\gamma\mathcal{D}_{\sigma_R}\rho(t)+\gamma\mathcal{D}_{\sigma_L}\rho(t)
\end{equation*}

where $\gamma=1/T_1$ is the Purcell-enhanced decay rate of the trion state, $\sigma_{R}=\ket{\uparrow}\bra{\downarrow\uparrow\Uparrow}$ ($\sigma_{L}=\ket{\downarrow}\bra{\downarrow\uparrow\Downarrow}$) is the optical lowering operator coupled to right (left) circularly-polarized light and $\mathcal{D}_\sigma\rho=\sigma\rho\sigma^\dagger - \sigma^\dagger\sigma\rho/2-\rho\sigma^\dagger\sigma/2$.

Lastly we do not explicitly model the time-dependent dynamic of excitation pulses, considering them as instantaneous and linearly polarized $\pi$-pulses, equivalent to the application of the following unitary rotation on the spin state:

\begin{equation*}
    R_\text{e}(\theta_{e})=\exp\left(-i\pi(\cos(\theta_{e})\sigma_{y,H}+\sin(\theta_{e})\sigma_{y,V})/2\right),
\end{equation*}
where $\theta_{e}$ is the angle between the polarization plane and the [110] sample axis, $\sigma_{y,H}=-i(\sigma_{H}-\sigma_{H}^\dagger)$, $\sigma_{y,V}=-i(\sigma_{V}-\sigma_{V}^\dagger)$ for $\sigma_H=(\sigma_L+\sigma_R)/\sqrt{2}$ and $\sigma_V=-i(\sigma_L-\sigma_R)/\sqrt{2}$. The exact value of the equation parameters used in the simulation are shown in Table \ref{tab:params}.

\begin{table}[htb]
    \centering
    \begin{tabular}{c|c}
         & \\
        \hline
        $B$ [$mT$] & 60 \\
        $T_1$[$ns$] &  0.2\\
        $g_e$ & 0.4 \\
        $q$ & -0.012 \\
        $\rho$ &  0.086\\ 
        $q_c$ &  -0.018\\ 
        $\theta_0$ [°] & -28\\
        $\theta_B$ [°] & [$0,360$] \\
        $\theta_{e}$ [°] & [$0,360$] \\
    \end{tabular}
    \caption{Value of the system and protocol parameters used to simulate the master equation of the spin-photon system}
    \label{tab:params}
\end{table}

This simulation framework allows one to set the relevant system and protocol parameter and then track the joint evolution of the spin-photon system under an arbitrary sequence of excitation pulses, photon detection and waiting times.

All simulations start from the completely mixed state $\rho_0 = \frac{1}{2}\ket{\uparrow}\bra{\uparrow} + \frac{1}{2}\ket{\downarrow}\bra{\downarrow}$ and the spin preparation step is handled explicitly as triggered emission plus a following polarization measurement.
Furthermore when precession and emission timescales are comparable the triggered emission of photons were observed to produce a delay in the subsequent precession of the spin state. For this reason, for every given set of simulation parameter, target $\pi/2$- and $\pi$-rotations were respectively optimized by algorithmic maximization/minimization of the simulated population difference.

The concurrence landscapes shown in Figures \ref{fig:fid}a and \ref{fig:landscape} were obtained by explicitly computing the figure of merit on the reduced density matrix obtained by simulating the optimized spin-photon entanglement protocol, as a function of the two protocol parameters $\theta_B$ and $\theta_{e}$.
The qualitative behaviour of the system depends only on the ratio between $q$ and $\rho$. We show in Fig.~\ref{fig:landscape} the three limiting cases of $\rho > q$, $\rho = q$ and $\rho < q$. The specific form of the $g$-factor thus has a large impact on the geometry of the configurations that allow optimal concurrence. 

Each point in Figure \ref{fig:fid}b comes from the simulation of realistic linear cluster state, computed by first fixing the values of $\theta_B$ and $\theta_{e}$ to one of the four selected regimes, then simulating explicitly the cluster state protocol with the optimized rotation times. In parallel we computed the ideal reference state, considering perfect state preparation, ideal spin rotations and precessionless excitation plus photon emission. The fidelity value quoted in Figure \ref{fig:fid}b was finally computed between these two density matrices.

\begin{figure}[htb]
    \centering
    \includegraphics[width = \linewidth]{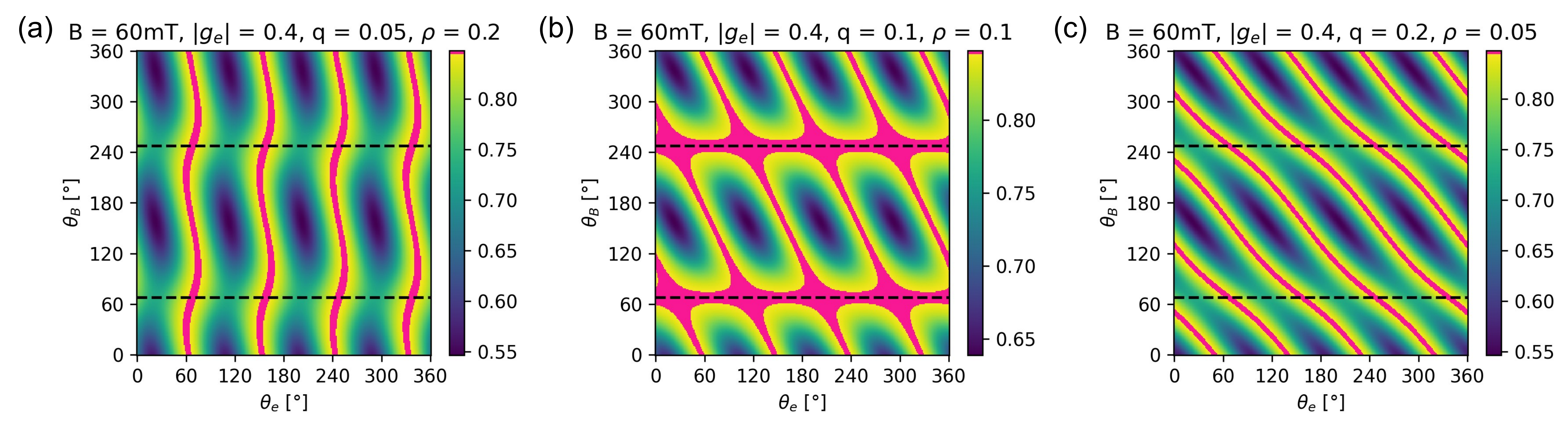}
    \caption{Concurrence of the spin-photon  produced by the entanglement protocol as a function of the angle of linear polarization of excitation $\theta_e$, and sample-magnetic field angle $\theta_B$, computed for different regimes of hole $g$-factor anisotropy: (a) $\rho > q$, (b) $\rho = q$, (c) $\rho < q$. For all three plots $q_c =0$ and $2\theta_0 = -\frac{\pi}{4}$. The red regions represent maximum entanglement, achieved when $\theta_e$ enables optimal addressing of the hole eigenstate. The dashed lines indicate the configurations where $\abs{g_h}$ (and accordingly hole state precession $\omega_h$) is minimized. The regime $\rho= q$ allows two values of $\theta_B$ which produce $\abs{g_{h}} = 0$, expanding the range of optimum concurrence.}
    \label{fig:landscape}
\end{figure}

\section{\label{sec:setup}Experimental setup}

\begin{figure*}[h]
\includegraphics[scale=0.1]{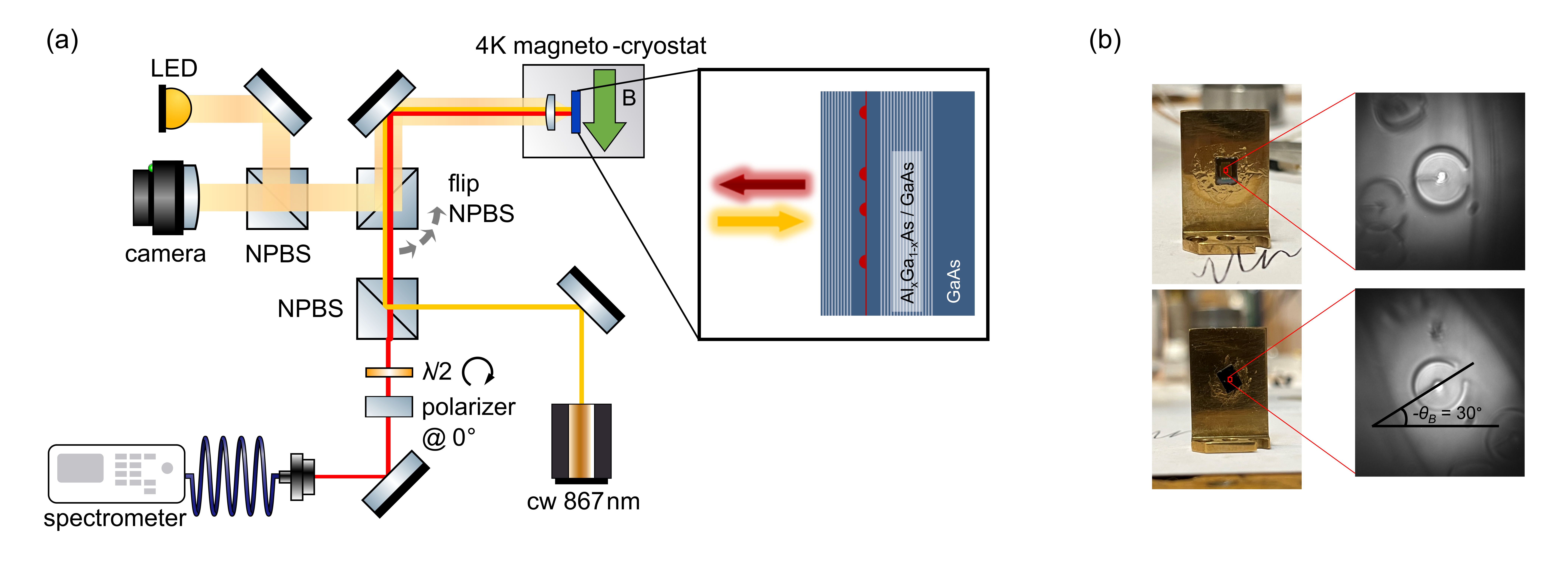}
\caption{\label{fig:exp}a) Top-down schematic view of experimental setup used to perform polarization-resolved photoluminescence measurements at 4K with a transverse magnetic field. NPBS = non-polarizing beamsplitter. b) Photos of the sample mounted at different angles $\theta_B$ with respect to the magnetic field, and corresponding images of the lithographically marked QD and angle measurements}
\end{figure*}

The complete experimental setup is shown in Fig~\ref{fig:exp}a. The sample is mounted in a 4K magneto-cryostat from MyCryoFirm. The aspheric lens (0.68~N.A.) used to focus on the sample is fixed inside the cryostat. The optical axis is parallel to the optical table. Inside the cryostat sits a split-coil superconducting magnet from CryoMagnetics fixed in Voigt geometry with respect to the sample optical axis. The sample is excited with a continuous wave tunable Ti:Sapphire laser from Spectra-Physics, set at 867~nm for this experiment. In the detection pathway there sits a polarizer  and an achromatic half wave-plate mounted on a motorized rotation mount. Photoluminescence spectra are collected with a dual-grating spectrometer providing 18~\textmu eV spectral resolution and a Princeton Instruments CCD Nitrogen-cooled camera. The lithographically-marked individual QDs are imaged using a standard CCD camera and NIR illumination by a LED. The images of the marked QDs shown in Fig~\ref{fig:exp}b are measured with ImageJ to identify the angle of rotation $-\theta_B$ with respect to the sample edge, equivalent to the real-space $0\degree$ of the magnetic field vector.
\end{widetext}

\end{document}